# Adaptive Virtual Neuroarchitecture

Abhinandan Jain, Pattie Maes, and Misha Sra

**Abstract** Our surrounding environment impacts our cognitive-emotional processes on a daily basis and shapes our physical, psychological and social wellbeing. Although the effects of the built environment on our psycho-physiological processes are well studied, virtual environment design with a potentially similar impact on the user, has received limited attention. Based on the influence of space design on a user and combining that with the dynamic affordances of virtual spaces, we present the idea of *adaptive virtual neuroarchitecture (AVN)*, where virtual environments respond to the user and the user's real world context while simultaneously influencing them both in realtime. To show how AVN has been explored in current research, we present a sampling of recent work that demonstrates reciprocal relationships using physical affordances (space, objects), the user's state (physiological, cognitive, emotional), and the virtual world used in the design of novel virtual reality experiences. We believe AVN has the potential to help us learn how to design spaces and environments that can enhance the wellbeing of their inhabitants.

## 1 Introduction

Twentieth century philosopher Maurice Merleau-Ponty wrote, space is existential and existence is spatial [46]. That is to say, we are always somewhere. Everyday we awake in space, move in space, eat, sleep and work in space. As Winnicott and Merleau-Ponty stated, our bodies are fundamentally related to the space that we

Abhinandan Jain
MIT Media Lab, e-mail: abyjain@mit.edu

Pattie Maes
MIT Media Lab, e-mail: pattie@media.mit.edu

Misha Sra
UCSB, e-mail: sra@cs.ucsb.edu

inhabit and this embodiment enables us to experience our lived reality, allowing us to not only perceive our own existence but also the world around us [29]. Though Murray [51] describes the body, as both, a "holistic sense organ " and an "assemblage of sense apparatus", we do not actively reflect on bodily perception in our daily lives. Despite non-conscious reflection on bodily perception by individuals, architects take the aggregate perceptual processes into account for designing the built environment. Knowing that people spend most of their daily time indoors, buildings can be an important focal point for promoting wellbeing by optimizing the occupant's experience of the indoors space. We believe bodily perceptual processes can similarly form the basis of built environment design in virtual reality (VR).

Though our interpretation of the real world is a continually evolving process through our interactions with the environment, the built environment itself is largely static. Most physical space adaptations are usually achieved through landscape or interior design. For example, urban planners create and maintain green spaces (e.g., parks, yards, terraces), personalizing them with thought given to things like visual impact, shade or other benefits (e.g., fruit, smell) and changes in color (e.g., fall foliage, annuals vs perennials) over time. Similarly, people expend considerable effort and time in designing their homes, choosing furniture, flooring, lighting, paint and colors to adapt their living spaces to themselves.

In recent years there has been "convergent agreement from architects, designers, psychologists, and neuroscientists about the multifactorial nature of the reciprocal interaction between humans and built space," and the desire to understand how it can impact human wellbeing [14]. Neuroarchitecture is an emerging term defined as the application of neuroscience specifically pertaining to the use of knowledge on human perception, cognition and behavior for the design of physical spaces [47]. It is a new interdisciplinary area that studies the interaction between the brain and built spaces [15]. We believe virtual neuroarchitecture is a natural extension of neuroarchitecture especially since VR experiences evoke a sense of presence or "being there," much like physical spaces do and lead to a different immersive experience than that of 3D video game worlds on a computer screen. In contrast with physical spaces where material, objects, effort, cost and time are required to modify and adapt the space to specific needs or purposes, virtual spaces can respond to the user and adapt to their cognitive and emotional states dynamically with an immediacy and convenience that is only possible with digital content. We call this **adaptive** (dynamic and reciprocal) connection between a physical space, a virtual space and the user, *adaptive virtual neuroarchitecture (AVN)*. By dynamic we refer to the automatic transformation of the virtual space in response to changes in the user's physical space and changes in the user themselves (e.g., physiological data, emotional data). By reciprocal, we mean the establishment of a continuous feedback loop between the virtual space, the physical space, and the user. Both types of connections eventually establish computer-to-human interaction, where a digitally generated world interacts with user's perception with the goal of influencing their cognitive and emotional states. In the design space of virtual neuroarchitecture, the example projects we present in Section 3.1 and 3.2 specifically focus on adaptive virtual environments. A virtual environment is composed of both inanimate and animate elements such as buildings, trees, furniture, animals and virtual humans. Each of these elements can respond to a user, either through direct interaction or automatically based on user data. For this work, we constrain our examples to changes in the inanimate elements that happen either via user interaction or automatically. Our hope is that this design space and the questions raised by the example projects highlight new opportunities for the design of computer-to-human interactions and for probing the potential impact of AVN on human wellbeing at physical, cognitive, emotional, and social levels.

We begin below by discussing embodied cognition, specifically how the mind and the body interacting together in an environment shape our actions and behaviors. To support this discussion we provide examples of the built environment's impact on an individual's affective and cognitive processes. In Section 3, we present AVN in more detail with a few example projects that utilize characteristics of the real world and the human body to create dynamic spatial experiences that have the potential to influence a user's affective and behavioral processes. More specifically, we explore the dynamics in two broad categories of *Adaptive Spaces* and *Adaptive Experiences*. Lastly, we highlight potential applications of AVN in education, training, wellbeing, and architectural design.

## 2 The Mind, The Body and The Environment

### 2.1 Embodied Cognition

In the last few decades, research has shifted from an understanding of the mind as an independent information processor towards understanding cognitive processes in relation to the environment [28]. This shift from Descartes' mind-body dualism towards embodied cognition has placed emphasis on the body and the environment in providing a foundation for facilitating and supporting cognition. Proponents argue that cognition takes place in the context of a physical environment, and continuously evolves through interaction with this environment which involves the processes of perception and action [80]. That is to say, the mind works collectively with the body towards making sense of the world. The process involves "sensation – gathering of information; perception – analysis of gathered information; cognition – synthesis of the information; and generalization – representation of the information" [39, 33].

Research suggests that not all incoming sensations and corresponding processing becomes part of conscious awareness. These sensations can factor into our unconscious (not aware), conscious (aware) or metacognitive (aware of being aware) processing [60]. Which pathway is used depends upon the signal's characteristics such as its attentional or cognitive demands or its correlation to prior mental representations [33]. Human cognition, affect or behavior can, thus, be influenced through multiple sensory processing pathways.

The non-conscious processing of information leads the mind and body into structuring actions even before awareness of intention to act forms. Such non-conscious prenoetic processes create the schema for potential actions within the environment the individual is situated in [8]. For example, the decision to grasp an object is determined by the presence of an object within reach and by our having hands. Similarly, seeing always happens from somewhere and it depends on the body and head pose of the viewer as well as any constraints in the surrounding environment [8, 64]. As Leder summarizes it, "To see something as reachable and thereby open to my use is to implicitly experience my body's capacity of reach" [37]. This implies that incoming sensory stimuli to the body are organized within the body schema and continuously updated with sensory input and body movement.This embodied model of processing and cognition applies to VR and the design of adaptive VR spaces, regardless of an avatar embodiment. A change in the user's situational space, i.e., the physical space the user is located in at the time of the VR experience, generates a new schema for potential actions and outcomes in the virtual space. We discuss a VR project that exemplifies this idea in 3.1.1.

## 2.2 Environmental Interactions with Psycho-physiological Processes

The view of cognition being embodied is consolidated with established knowledge of the effects of built space on perception, affect, and behavioral responses [76]. "We shape our buildings; thereafter they shape us" said Sir Winston Churchill in his speech on the rebuilding of the House of Commons, emphasizing the impact of shape (oblong instead of semi-circular) and size (not big enough to contain all Members) on governance. Churchill understood the influence of the built environment on its inhabitants. Research has since shown how a building's characteristics such as shape, lighting, acoustics, texture, colors or scale, influence the thoughts, emotional states, cognitive load, health, and behavior of its occupants [24, 38, 56, 17]. Our surrounding environment is an essential part of our daily lived experience with many people spending over 90% of their day inside a building. Therefore, indoors spaces have continuous cognitive-emotional influence on shaping our behavior [15].

### 2.2.1 Perceiving Space Through the Body

Perceiving space involves sensory input into the eyes with head and body movements determining what the eyes look at. Different types of body movements allow us to apprehend proximal and distal spaces through head turning or object grasping and translation, respectively. Humans experience space at three main scales of near, middle and far [27]. Near allows us to understand small object geometries by grasping and manipulating them with our hands. Middle space is experienced at a room-scale where lighting, contrast, and texture become important in helping understand the objects, the space and its relationship to the body. In middle space, only portions of objects are visible at any moment when viewed from a specific location. However, the user can easily move around the space to interact with objects e.g., a user can walk up to a table or sit down on a couch. At the far scale, touch-based interaction is not possible and our ability to perceive objects is limited to understanding forms and shapes as viewed from a distance (e.g., the outlines of buildings or shapes of trees). Complexity and details that are perceivable at near and middle scales are often not perceivable at this scale. Perception of space, similar to the middle space, is influenced by lighting and texture but also by shadows, vistas, skylines and the ability to move across longer distances, whether by foot or in a vehicle. We situate the study of virtual neuroarchitecture in the middle and far spaces where the environment's features, both spatial and object-based, can start to interact and shape cognitive processes (Figure 1). VR allows us to design experiences in near, middle and far spaces, in contrast to AR which, in practice, is largely limited to near and middle space experiences. While object-based interactions in any physical environment are limited to near and middle spaces, in a virtual environment even far spaces can enable object interaction through digital enhancements of a user's abilities such as Gravity Gloves as seen in the Half-Life:Alyx VR game which allow a user to pull objects to their hands or virtual arm lengthening to reach far objects [55] or teleportation to allow rapid movement across space [10].

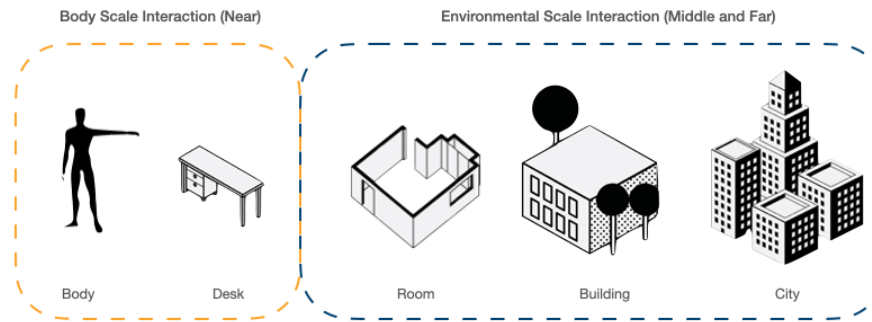

**Fig. 1** Near spaces are within reach of the body and we categorize them as body scale spaces. Middle and far spaces are where the surrounding environment and architecture starts impacting us.

### 2.2.2 Space and Wellbeing

Architects and urban designers have employed all levels of scale and the body's relationship to space in the design of cities, buildings, public (indoors and outdoors) and private spaces for centuries. Prior work has demonstrated the impact of such architectural and environmental characteristics on cognitive processes. For example, static properties such as materials, colors and dynamic elements of space such as background noise have been shown to facilitate individual creative performance [43, 45]. Correspondingly, exposure to poor architectural and indoor environmental conditions have been found to negatively impact occupants' cognitive functions such as decision-making [2, 42], and shown to have adverse effects on performance and productivity [26, 81]. Environmental psychology research has helped design hospi-tals, schools, and residential units that can lead to optimal cognitive performance conditions or increase productivity. For example, studies have found that in classrooms with increased natural light, students achieved higher test scores in contrast to those with lower natural light [50]. For spaces where access to views of the outdoors are not possible, studies have shown simulated green settings to have positive effects on mood, self-esteem and self-reported feelings of stress and depression [9, 5, 44]. Prior work has also shown how the addition of "hospital green" walls and nature landscape views from windows helped speed up the healing process of patients by 8.5% [77] and reduce need for pain relief [74]. Given that humans in urban areas tend to spend inordinately large amounts of their time indoors, even more so given the recent Covid-19 pandemic, the design of space and understanding its impact on our wellbeing have never been more critical.

### 2.2.3 Multisensory Experiences in Space

Multisensory stimulation and synesthetic experiences have been found to be a causal factor for generating aesthetic emotions and creating metanoic (spiritually moving) or transformative experiences for the observer in an architectural space [65]. For example, the scale, lighting and synchronous reverberation in a cathedral create an immersive and almost hypnotic multi-sensory experience with audio, visual and emotional input [7]. However, incongruency of such stimuli can create psychological tension and cognitive conflict. As shown in prior work, disproportionate spatial features can lead to strange sound reverberations causing discomfort or the poor use of colors and lighting can impact an occupant's mood negatively [7].

Just as we respond cognitively and emotionally to the built environment, as shown in the examples presented above, virtual environment properties, such as color, lightning, and texture, also impact us akin to the impact felt in the real world [52]. Additionally, dynamic elements of the physical environment which adapt to the user have been shown to enhance a user's cognitive processes [84, 16], mood [53], and wellbeing [23]. For e.g. Mediated Atmosphere uses a digital display to change the perceived environment based on the user's physiology and shows positive impact on the user's focus and stress levels [84]. Such demonstrated effects, of static and adaptive elements in the physical environment, on a user's psycho-physiological processes is our primary inspiration to explore the design of adaptive (dynamically responsive and reciprocal) virtual environments.

## 3 Adaptive Virtual Neuroarchitecture

Architecture has provided us with the built environment, surrounding us in structures which integrate elements that can influence our physiological, psychological and social-cognitive processes [75, 38]. Neuroarchitecture is an emerging interdis-ciplinary area that applies the understanding of perception, cognition and behavior from neuroscience to architectural design with the goal of optimizing the positive impact of the built environment on the occupants at cognitive-emotional and behavioral levels [15, 31]. However, the physical world is largely static and renovating the physical environment frequently to study, evaluate or optimize the interaction between the brain and built spaces and subsequent influence on a user's cognitive processes can be expensive, cumbersome and logistically less feasible. Recent work in the design of flexible architectural spaces explores using robotic elements to reconfigure the user's space at the press of a button. While this makes the physical space dynamic, it may be limited in functionality due to the modifications being constrained by the physical space extents [36].

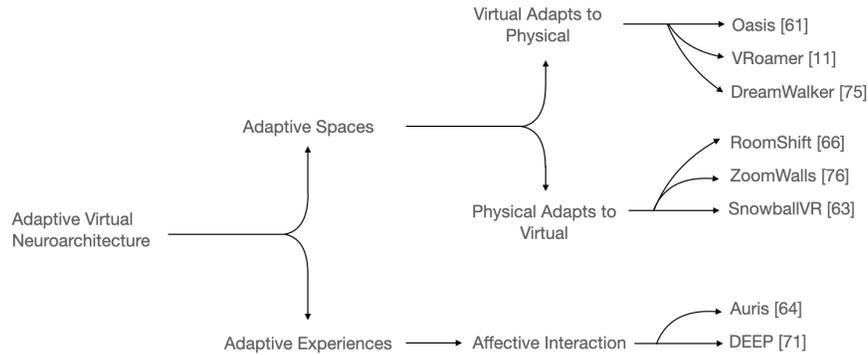

**Fig. 2** We categorize adaptive virtual neuroarchitecture (AVN) into Adaptive Spaces and Adaptive Experiences and showcase example projects in each category.

VR presents opportunities for designing architectural spaces with a layer of dynamism which can allow us to achieve, in a sense, "programmable matter." The term "programmable matter" was introduced in 1991 by Toffoli and Margolus [1] to refer to matter that has the ability to change its physical properties.

With AVN, we propose the design of VR spaces and experiences that are dynamic, responsive and integrate reciprocity between the physical and virtual environments and a user's physiological and cognitive states. We can map a physical environment's properties such as texture, color, lighting, acoustics, size, layout or style to the design of a virtual environment and modify them often and as quickly as desired. Furthermore, we can map a user's physiological or psychological responses to virtual space modifications and create feedback loops between the virtual dynamism and the physiological signals to create a sense of harmony between the two. The dynamic remodeling of the virtual environment can be frequent, continuous or occasional depending on what the user needs. The remodelled virtual environment can translate across different scales (near, middle or far) depending upon the user's situation and desire. The degree of virtual responsiveness can be manipulated by the degree of input signal (user's sensor data and/or changes in their physical space) while the direction of change can be determined by the user. For example, if the user's heart rate increases slowly, the environment can respond to it slowly with modifications to features such as lighting, color or space. These modifications can either be in a direction that results in an increase in the user's heart rate or calms them down, which could be decided by the user or the interaction designer. However, allowing developers to make the decision on how to modify a user's physiological and psychological responses raises ethical concerns related to limiting user agency and manipulating the user to a degree potentially greater than achievable by changes in visual imagery alone. We can also create multi-sensory feedback loops that include stimuli other than the visual and auditory signals such as olfaction [4], proprioception [68, 41], and temperature [11]. These multi-sensory feedback loops can each uniquely impact cognitive processes (e.g., healing, mindfulness, wellbeing) by augmenting changes to the virtual environment and reciprocal changes to the user's perception of the updated virtual environment. We can imagine the virtual environment to be a fluid or liquid space that continuously morphs due to the user's presence and movement through it. Multi-sensory haptic feedback such as electrical muscle stimulation [40], weight perception [18], and terrain texture [34] can further increase the user's sense of presence in the VR space and reinforce desired environmental effects on the user through simultaneous changes in elements such as lighting, weather, and atmosphere.

---

[1] https://en.wikipedia.org/wiki/Programmable_matter

To better understand AVN, we have designed and built a few working prototypes, some of which we present here along with work by other researchers. We present these example projects in two broad categories of *Adaptive Spaces* and *Adaptive Experiences* (Figure 2). In *Adaptive Spaces*, the virtual and the physical world interact with each other, with the virtual environment often adapting to the physical space and in some cases, the physical space changing in response to the virtual space. In *Adaptive Experiences* we present VR projects which build upon physiological or psychological constructs that influence our cognitive processes (Figure 2). This not a comprehensive survey of prior work. Our primary goal is to highlight AVN in research and use the selected examples to present new and interesting design opportunities for future work in this area.

### 3.1 Adaptive Spaces

One of the major challenges in the design of adaptive VR environments is the mapping of a physical space to a virtual space such that the user is able to take advantage of the real world spatial and tactile affordances which can augment their virtual experience. Most VR systems render a pre-defined tracking space boundary to separate the physical world from the virtual, breaking the user's experience at the boundary by a visual or haptic reminder of the divide between the two seemingly separate worlds anytime the user approaches the boundary. While this is understandably done to prevent the user from colliding with physical objects, there is opportunity for more advanced integration between the user, their physical environment and the virtual world.

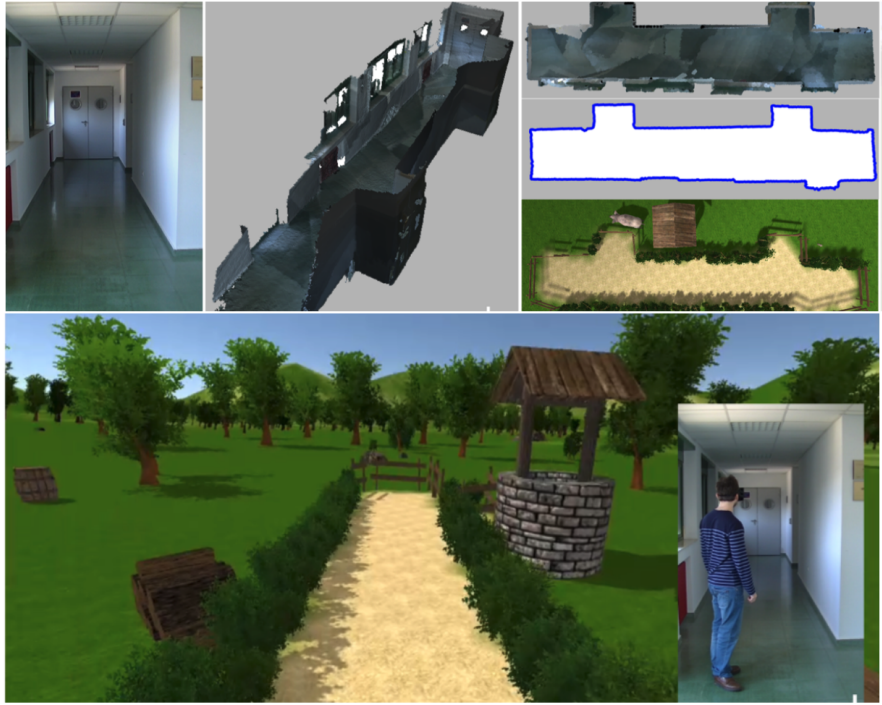

**Fig. 3** The Oasis system creates a 3D map of the physical environment, automatically detecting open walkable areas which allow the system to generate virtual environments with corresponding walkable areas. While the physical affordance of walking is the same in both environments, the visual experience and spatial feel can be quite different. For example, as seen in the figure above, a hallway is transformed into an open countryside which still allows the user to walk freely while wearing a wireless headset device in the mapped physical/virtual space without breaking immersion. Figure from the paper [67].

One option that has been explored in generating virtual spaces that adapt to the user's physical environment is using a depth camera to scan the user's surroundings, reconstruct geometry, infer walkable areas, and detect physical objects. This data is used to generate a VR environment which replaces the reconstructed walls and furniture with similarly sized and positioned virtual elements such that the user can freely walk in the physical space without running into any obstacles. The objects in the physical space provide haptic feedback to the user through their emulated virtual counterparts. [67, 61, 63]. A related exploration has used robotic or human manipulated furniture or walls that conform to changes in the virtual environment to provide the user a corresponding adaptable physical space and related haptic feedback [73, 83]. Using the real world as a template, a single physical space can be used to create multiple unique and visually different virtual environments with tactile feedback in ways that maintain user safety by preserving the geometry and semantics of the real world [67, 61]. The idea can be expanded into creating blended spaces for remotely located multi-user experiences that allow each user to freely walk around and interact as if they are in the same physical space [66]. Exploring the degrees to which the physical world is blended with a virtual world [30] can help understand how we perceive these hybrid spaces, paving the way for a future where the virtual and the physical will allow for seamless transition between them, without risk of injury.

### 3.1.1 Virtual Adapts to Physical

**Oasis**

Oasis automatically generates immersive and interactive VR worlds that adapt to the user's physical space [66, 67]. The system uses a 3D capture of an indoor scene to isolate walkable areas, interactions and obstacles. Individual objects from the physical world (e.g., chairs) are mapped to objects in the virtual environment and tracked in real-time for providing haptic feedback (Figure 3). The mapping of physical objects to their virtual counterparts creates a dynamic experience where users receive full haptic feedback from their interactions with virtual proxies. The user's actions in the virtual environment are also reflected in the real world (e.g., sitting down in a virtual/physical chair) and chair movements in the physical world are reflected in the corresponding movements of the virtual chair. This dynamic (realtime response) and reciprocal (bidirectional) relationship creates a connected situated experience between the physical and the virtual world. This subsequently increases presence for the user in the virtual world and creates a more engaging user experience. Oasis presents a step towards creating fully dynamic relationships between the physical and virtual worlds. Oasis also highlights transformation of space scale from middle space in the physical world to a far scale in the virtual world while mapping characteristics of the physical world dynamically to the virtual world. However, Oasis has limited object tracking and is bounded by finite walkable areas of the real world environment. It therefore does not exemplify a strong reciprocal spatial relationship where all changes in the virtual world are reflected in the physical world.

**VRoamer and Dreamwalker**

VRoamer [13] is a proof of concept system which detects collisions with obstacles in realtime and places virtual obstacles in the user's path to prevent them from a potential collision (Figure 5). VRoamer builds upon Oasis but differs in that the collisions with physical objects are dynamically detected and represented in the virtual world when walking on a pre-planned indoor path. Dreamwalker [82] builds upon VRoamer where it improves tracking while walking outdoors wearing a VR headset and includes walkable areas such as slopes, curbs and non-planar surfaces. It uses GPS data and SLAM for tracking and places a pre-created virtual environment along a known physical path for an outdoor walking experience in VR. Dreamwalker allows for longer walkable paths with realtime collision detection of moving obstacles (Figure 4). Whereas in Oasis a movable object like a chair is pre-scanned and only tracked in realtime, Dreamwalker allows for detecting previously unseen obstacles and representing them in the VR environment at runtime. In both VRoamer and Dreamwalker, representations of detected obstacles (static and moving) create a dynamic connection between the physical and virtual worlds as interactions in the physical space are reflected in the virtual space. However, neither VRoamer nor Dreamwalker establishes a reciprocal relationship between the real and virtual spaces since changes in the virtual world are not reflected in the physical environment.

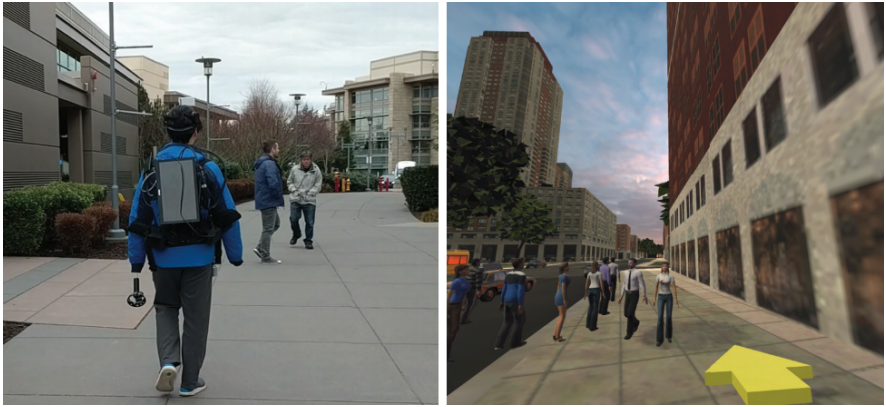

**Fig. 4** Dreamwalker creates a real-time VR experience in an outdoor physical environment. Figure from the paper [82].

Oasis, VRoamer and Dreamwalker are examples of virtual neuroarchitecture systems that enable multiple visual reconfigurations of the user's static physical space virtually, while maintaining a connection between the two spaces based on the affordance of walkability. Adaptive virtual spaces open up the opportunity to study the impact of size (closed room vs open yard), shape (smooth curves vs sharp edges and corners), textures, colors, lightning, or weather on the user while maintaining the walkability of the user's actual environment. The automated generation in Oasis and real-time mapping in VRoamer combined with the outdoors path planning in Dreamwalker may one day allow for building VR environments on a city wide scale which adapt to create personalized experiences for all individuals.

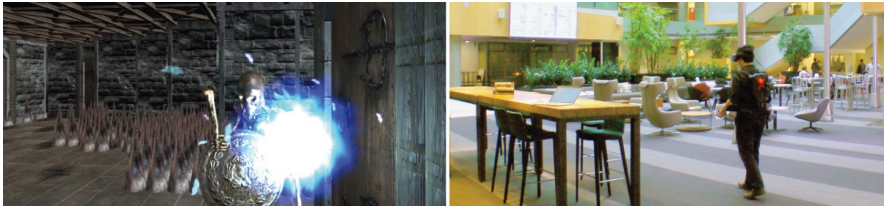

**Fig. 5** VRoamer's dungeon environment with virtual obstacles shown in response to physical obstacles in the user's environment. Figure from the paper [13].

### 3.1.2 Physical Adapts to Virtual

**RoomShift and ZoomWalls**

RoomShift [73] is a room size dynamic environment where physical objects move in response to a user's virtual movements and position themselves in appropriate locations around the user to provide haptic feedback (Figure 6). It utilizes a swarm of shape changing robotic assistants, similar to shelf-moving robots. The goal is to enable physical interactions such as touching, sitting, placing things and leaning while the user is immersed in VR. The system continuously tracks virtual touchable surfaces in close proximity to the user in VR and directs the robot swarm to transport physical objects to the positions without clashing with other robots or the user.

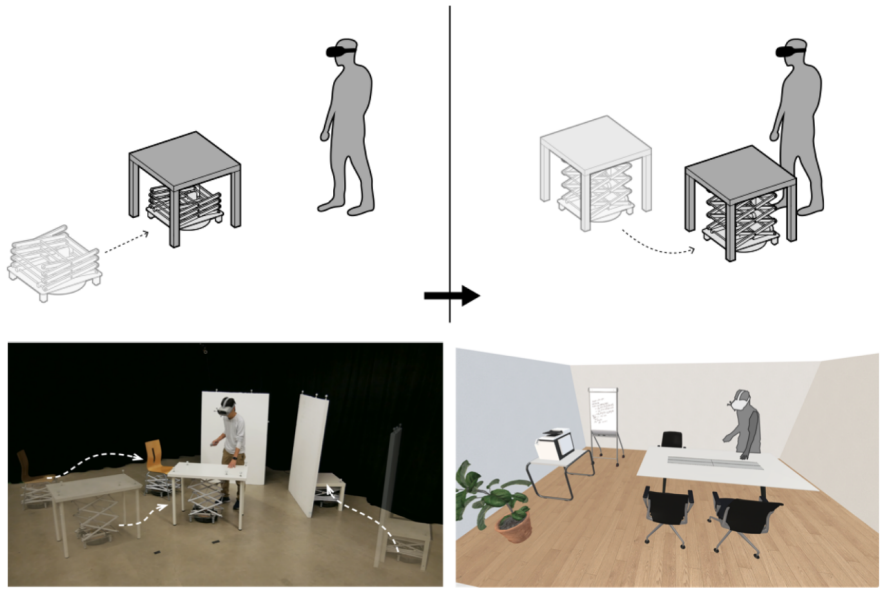

**Fig. 6** In RoomShift, robotic assistants move furniture around the user who is immersed in VR, demonstrating a way to represent changes in the virtual environment on the user's physical space. Figure from the paper [73].

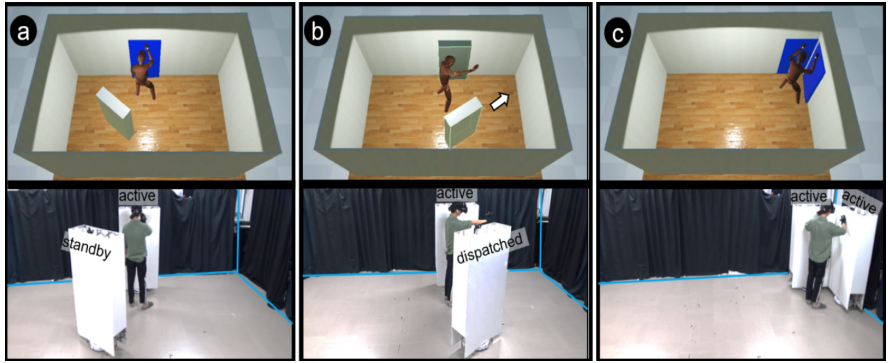

**Fig. 7** ZoomWalls showcases movement and organization of robotic walls around the user in response to changes in the virtual environment. Figure from the paper [83].

ZoomWalls [83], similar to RoomShift, creates a dynamic physical environment by reconfiguring robotic props that match interaction with virtual structures like walls/doors. The robots re-position themselves so as to create a simulated wall segment of the room/area (Figure 7). This simulation allows to create size-based illusions of the surroundings for the user. Having walls follow closely makes the user feel they are enclosed in a much tighter space while distributing them makes the space size seem bigger. The user study demonstrates increase in presence and immersion for the users.

**SnowballVR**

A dynamic match between a fort made of actuated cardboard boxes with a corresponding virtual snow fort enables manipulating physical objects based on virtual events in SnowballVR [69]. The snow fort takes advantage of the connection between the virtual and physical to provide haptic (boxes), aural (sound of boxes collapsing) and spatial (matching snow fort in position, scale and orientation) feedback to the user. Additionally, a visual mechanism that controls the collapsible boxes can engage the non-VR users (Figure 8) in this spectator focused VR experience. Without the need for a screen, non-VR users can determine how the virtual snowball fight is progressing based on how many snow fort boxes they see and hear collapse. The connection between the virtual and the physical is unidirectional, from VR to the physical world.

RoomShift, ZoomWalls and SnowballVR showcase the dynamic modulation of physical space in response to the user's behavior in a virtual environment, and do so without pre-configuration, as has been necessary in prior work exploring passive haptics for room-scale feedback [32].These projects show both a uni- and bi-directional relationship between the virtual and physical spaces as objects are updated in both worlds when their properties change in either. This opens up new questions related to boundaries and evolution of a user's sense of presence and being in a space when space itself becomes fluid.

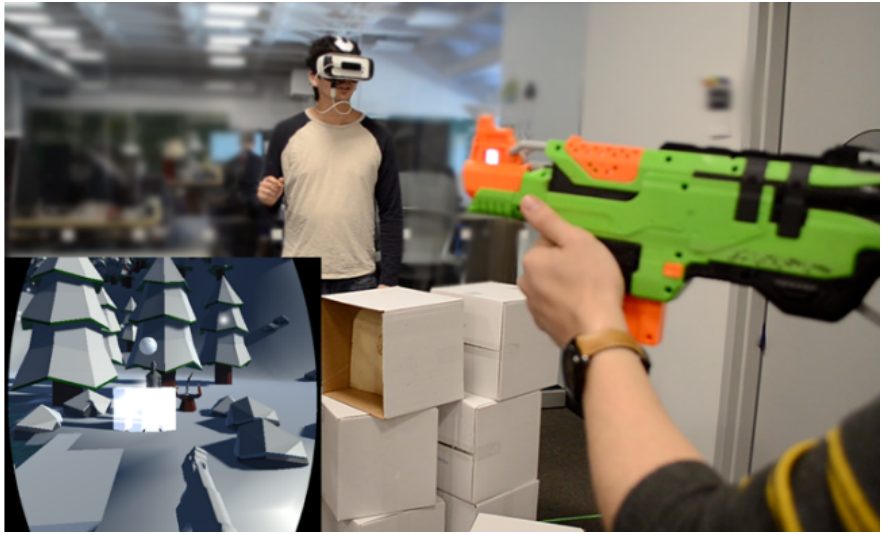

**Fig. 8** SnowballVR showing two users, one shooting virtual snowballs using a modified Nerf gun and the other dodging the snowballs by hiding behind a physical/virtual snow fort. Figure from the paper [69].

### 3.2 Adaptive Experiences

From the modest barn to the majestic monuments, the elements of architectural design like proportion and scale, light, color, and shape have been shown to impact humans in tangible ways. Whether it is entering the Cathédrale Notre-Dame de Paris for the first time and feeling simultaneously awed and humbled or entering a cavernous office daily and feeling time slow down, space has an undeniable emotional impact on both its short- and long-term inhabitants. Such psychological influence of architecture presents opportunities for virtual experiences that are personalized to the user for palpable cognitive and emotional impact. In this section we present two example projects where dynamic virtual experiences impact physiological and psychological states of the user.

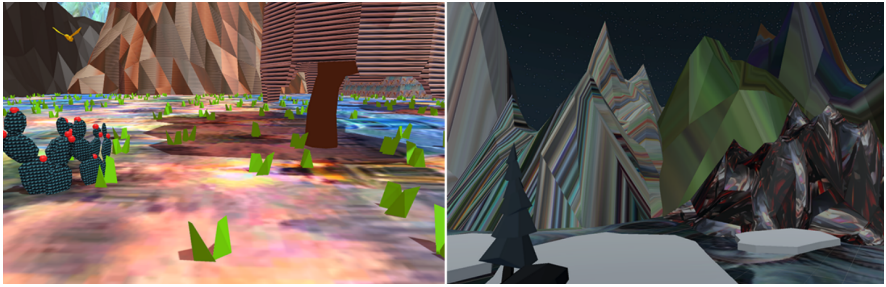

**Fig. 9** Auris generates mood-based VR worlds from music to create spatial and interactive experiences of listening to music. Happy (left) and sad (right) worlds generated from "The Bird and The Worm" and "Blue Prelude" by Nina Simone respectively. Figure from the paper [70].

### 3.2.1 Affective Interactions

**Auris**

Auris [70] is a VR world generation system that uses data from songs (audio and lyrics) to create virtual worlds that encapsulate and represent the mood of the song visually through space, texture, lighting and design. Layers of textures created using Inception ² with song data allow for the creation surreal environments for presenting a new type of dynamic and immersive 3D music experience (Figure 9). The user study showed that after experiencing the virtual environment, the user's mood reflected the mood depicted by the generated world. We would like to note that the choice of direction of influence, i.e., if the user is in a sad mood should they enter a happy environment or a sad environment, should reside with the user since there is possibility of making the user's mood worse by forcing a different mood on them. Based on this work, the multi-sensory feedback and interactive experience of the song-based space has the potential to impact a user's mood and can be used for the design of applications related to therapy, relaxation or stress management. Auris represents an example of dynamic environment transformation based on a detected mood in a song which is a psychological construct that influences the user's experience of and response to the environment. While a song's mood data is labeled collectively by a large group of individuals, the impact of the song-based VR space is evaluated on the individual.

**DEEP**

DEEP [78] is an immersive underwater fantasy VR world aimed at providing a calm and relaxing experience for the user (Figure 10). DEEP uses biofeedback to support breathing for a meditative experience. The system uses variable resistor stretch sensors to measure diaphragm expansions and modifies the virtual world using that data in real time. For example, based on the user's breathing cycles, the environment changes motion dynamics of objects resulting in a biofeedback loop. The goal is to use respiration physiology to help enable anxiety regulation and improve cognitive performance [78]. The feedback loop demonstrates a reciprocal relationship where the environment transforms and influences the user's physiology implicitly and vice-versa.

---

² https://ai.googleblog.com/2015/06/inceptionism-going-deeper-into-neural.html

Both Auris and DEEP showcase the use of psychological and physiological constructs respectively in the generation of a virtual world which in turn effects the user. Both projects present the design of virtual environments that are either based on or closely integrated with a user's physiological or psychological responses. While there is a lot of prior work that explores the integration of physiological signals with interaction in VR [22, 71], there are limited projects that specifically alter the virtual environment based on the user's data.

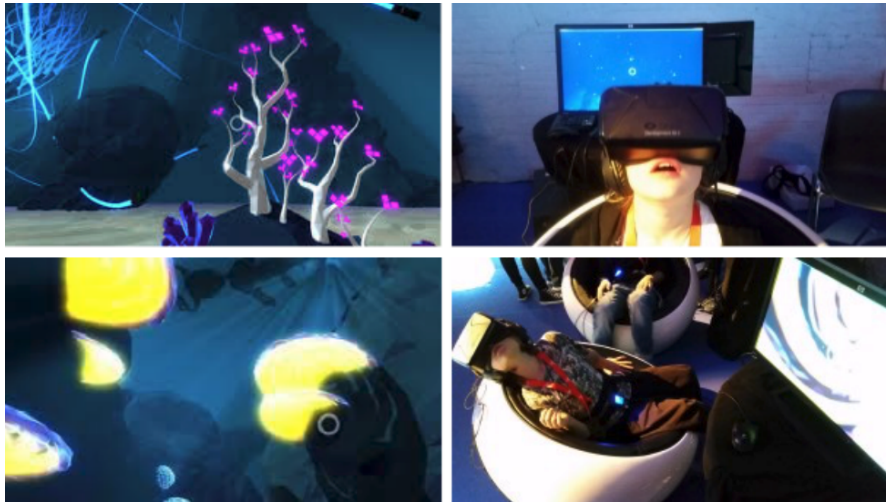

**Fig. 10** DEEP is an underwater virtual world that responds to the users. The figure shows user's experiencing it at the 2015 Cinekid Festival. Figure from the paper [78].

## 4 Applications

AVN presents a new design space for developing experiences which are intertwined with a user's surroundings, physiology or psychology. Such experiences allow for subtle or overt environmental (physical and virtual) manipulations which in turn have an effect on the user such as impacting cognitive functions and improving wellbeing. With newer devices like PhysioHMD (Galea[3]) [6], there is opportunity to simplify the process of conducting neuroarchitecture experiments. Here, we present a few potential applications of AVN in education and training, wellbeing, and architectural design which are largely inspired by broad areas of interest in the VR research community. Many other areas of interest in virtual, augmented and mixed reality (XR for short) such as entertainment, socialization, and teamwork could also benefit from the concept of AVN.

---
[3] https://galea.co/

### 4.1 Education and Training

Our bodily and environmental experiences are closely linked in facilitating the learning process [62, 72]. Researchers have emphasized the significance of being in and experiencing an environment in order to learn effectively [49]. Environmental characteristics such as illumination have shown positive effects on the learning process [50]. With AVN we imagine creating personalized and adaptive environments which optimize mood and emotions based on user's physiology, to facilitate learning. For example, changing environmental settings based on user's cognitive load and affective state to personalize the learning process, matching the pace and difficulty of content to the user. Another way researchers have shown the effectiveness on learning and memorization is by using spatially distributed cues in a user's surroundings. For example, NeverMind [57] uses the 'method of loci' memorization technique in augmented reality to help users memorize more effectively, creating a connection between a familiar walking path and corresponding virtual objects placed along that path. This inspires the idea of using learning materials by spatially distributing and embedding them in a virtual environment to influence learning motivation, engagement and recall.

VR has been used for training since the mid-90s with consumer VR devices now increasingly being used for medical [58], safety [59], customer experience[4] and other forms of training. Virtual training requires a close match between the physical space and the virtual worlds as well as a match in the actions performed in order for the experience to meaningfully transfer to real situations. By creating virtual environments, interactions and objects that are similar to their real world counterparts, with or without haptics, training scenarios have already proven to be better than video learning of similar tasks [1, 12].

With AVN, we believe, creating a continuum of experiences across multiple characteristics of virtual and physical worlds enables utilization of the virtual world reconfigurability with the physical world tangibility for the design of new types of dynamic learning and training experiences.

---

[4] https://www.strivr.com/blog/customer-experience-training-virtual-reality/

## 4.2 Therapy and Wellbeing

Research has shown various effects of the surroundings on an individual's psychological state such as mood, anxiety, stress [56, 9, 5, 7] as well as on healing processes of the body [74, 77]. A now famous study by Langer et al. [35] demonstrated positive impact of surroundings on the physical health and wellbeing of elderly subjects, when they retreated to a setting that was reminiscent of their 20 years younger selves. Virtual neuroarchitecture provides this potential of recreating time shifted environments based on surrounding physical environments as the examples in Section 3.1.1 show. Researchers have also used VR as an affective intervention to induce calmness and reduce anxiety by creating real-time adaptive environments based on the user's physiological signals [3, 25]. These projects are presented in Section 3.2.1. Adaptive experiences for therapy can be enhanced with wearable devices which may track data over time as well as provide vibrotactile feedback. Using false feedback techniques have also been demonstrated to increase focus and calmness [20, 21, 19] through the integration of the user's physiological data into their experience.

## 4.3 Architectural Design

VR environments are being utilized in CAAD (Computer Aided Architectural Design) for simulating indoor spaces before finalizing architectural designs [48]. Spatial and acoustics modeling of an environment is simulated before construction begins to get a better understanding of the characteristics of the space [79, 54], especially crucial for the design of music and other sound related venues but also valuable for the design of quiet spaces like churches or apartment buildings for maintaining privacy. Research has identified the need for more immersive, scalable, multi-sensory and interactive experiences as a future challenge for creating better CAAD tools [48, 54].

AVN has the potential to emulate dynamic user-based spatial modifications from the virtual environment to the physical environment as presented in Section 3.1.1. The ease of re-configurability from the virtual environment and concurrent reciprocity in the physical environment present an opportunity to understand the experience and impact of multiple designs in parallel, allowing for rapid modifications or granular adjustment of individual parameters, enabling evaluation of a space over time and and testing with as few or as many users as needed, before expending time and cost in construction.

## 5 Conclusion

Our surrounding environment impacts us continuously in our daily lives, affecting our physiological and psychological processes. Both indoor and outdoor spaces can impact our experiences and motivations and shape our behaviors. However, the built environment is often static, impersonal (especially shared indoor spaces and public outdoor spaces) and inflexible. VR as a tool addresses these shortcomings and presents new opportunities for transforming spaces into dynamic experiences with the potential to positively impact users.

In this chapter, we presented the idea of *adaptive virtual neuroarchitecture (AVN)* and defined it as a dynamic and reciprocal connection between the physical space, the virtual space and the user. AVN designs can not only adapt to and transform the user's surroundings but also their mental state, often through a feedback loop that integrates a user's physiological data into the virtual experience. Our idea of AVN builds on current knowledge of embodied cognition to inform the design of adaptive virtual environments as well as the emerging field of neuroarchitecture. Through a few example projects from our own work and others we presented how adaptive virtual worlds and experiences that interact with both the user's surroundings and their cognitive states can open up new research questions and opportunities for future work in the design of virtual environments and experiences.